\documentclass[aps,prb,twocolumn]{revtex4-1}
\usepackage{graphics}
\usepackage{subfigure}
\usepackage{longtable}
\usepackage{graphicx}
\usepackage{pstricks}
\usepackage{amsmath}
\usepackage{dcolumn}
\usepackage{bm}

\newcommand{\bra}[1]{\langle#1|}
\newcommand{\ket}[1]{|#1\rangle}

\begin{document}

\title{Near-degeneracy of extended $s + d_{x^2-y^2}$ and $d_{xy}$ order parameters in 
quasi-two-dimensional organic superconductors}

\author{Daniel Guterding}
\email{guterding@itp.uni-frankfurt.de}
\affiliation{Institut f\"ur Theoretische Physik, Goethe-Universit\"at 
Frankfurt, 
Max-von-Laue-Stra{\ss}e 1, 60438 Frankfurt am Main, Germany}

\author{Michaela Altmeyer}
\affiliation{Institut f\"ur Theoretische Physik, Goethe-Universit\"at 
Frankfurt, 
Max-von-Laue-Stra{\ss}e 1, 60438 Frankfurt am Main, Germany}

\author{Harald O. Jeschke}
\affiliation{Institut f\"ur Theoretische Physik, Goethe-Universit\"at 
Frankfurt, 
Max-von-Laue-Stra{\ss}e 1, 60438 Frankfurt am Main, Germany}

\author{Roser Valent\'i}
\affiliation{Institut f\"ur Theoretische Physik, Goethe-Universit\"at 
Frankfurt, 
Max-von-Laue-Stra{\ss}e 1, 60438 Frankfurt am Main, Germany}

\begin{abstract}
The symmetry of the superconducting order parameter in quasi-two-dimensional 
BEDT-TTF organic superconductors is a subject of ongoing debate. We report {\it 
ab initio} density functional theory calculations for a number of organic 
superconductors containing $\kappa$-type layers. Using projective Wannier 
functions we derive parameters of a common low-energy Hamiltonian based on 
individual BEDT-TTF molecular orbitals. In a random phase approximation 
spin-fluctuation approach we investigate the evolution of the superconducting 
pairing symmetry within this model and point out a phase-transition between 
extended $s + d_{x^2-y^2}$ and $d_{xy}$ symmetry. We discuss the origin of the 
mixed order parameter and the relation between the realistic molecule 
description and the widely used dimer approximation. Based on our {\it ab 
initio} calculations we position the investigated materials in the obtained 
molecule model phase diagram and simulate scanning tunneling spectroscopy 
experiments for selected cases. Our calculations show that many $\kappa$-type 
materials lie close to the phase transition line between the two pairing 
symmetry types found in our calculation, possibly explaining the multitude of 
contradictory experiments in this field.
\end{abstract}

\pacs{
  71.15.Mb, 
  71.20.Rv, 
  74.20.Pq, 
  74.70.Kn  
}

\maketitle

\section{Introduction}
Quasi-two-dimensional organic charge transfer salts of 
bis-ethylenedithio-tetrathiafulvalene molecules, abbreviated BEDT-TTF or ET, 
have attracted much interest due to their rich phase 
diagrams~\cite{Toyota2007, Powell2006}. Among these materials, the family of 
$\kappa$-(ET)$_2 X$ salts, where $\kappa$ refers to a specific arrangement of 
the ET molecules and $X$ corresponds to a monovalent anion, stands out due to 
the realization of fascinating states of matter like Mott insulator, 
unconventional superconductor 
or spin-liquid~\cite{Toyota2007, Powell2006, Elsinger2000, Shimizu2003, 
Kurosaki2005, Kagawa2005}. Especially the immediate vicinity of the 
superconducting phase to an antiferromagnetic Mott insulator suggests a deeper 
connection between two-dimensional organics and high-temperature cuprate 
superconductors~\cite{McKenzie1997}.

Although superconducting $\kappa$-type charge transfer salts have been 
investigated, for instance, in studies of specific heat~\cite{Elsinger2000, 
Mueller2002, Wosnitza2003, Taylor2007, Taylor2008, Malone2010}, surface 
impedance~\cite{Milbradt2013}, thermal conductivity~\cite{Izawa2001}, 
millimeter-wave transmission~\cite{Schrama1999}, scanning tunneling spectroscopy 
(STS)~\cite{Arai2001, Ichimura2008, Oka2015, Diehl2016} and elastic 
constants~\cite{Dion2009}, no consensus about the symmetry of the 
superconducting pairing has been reached so far. Some of the experiments are in 
favor of $s$-wave symmetry~\cite{Elsinger2000, Mueller2002, Wosnitza2003}, while 
other studies have proposed $d$-wave states with contradictory positions of the 
nodes in the superconducting order parameter~\cite{Taylor2007, Taylor2008, 
Malone2010, Milbradt2013, Izawa2001, Schrama1999, Arai2001, Ichimura2008, 
Oka2015}. Evidence for a mixed-symmetry order parameter was recently provided in 
Refs.~\onlinecite{Dion2009, Diehl2016}. Notably, evidence for a phase separation 
between different $d$-wave states has recently been reported in 
Ref.~\onlinecite{Oka2015}.

In theoretical approaches, the $\kappa$-(ET)$_2 X$ family of materials is often 
described by a half-filled Hubbard model of (ET)$_2$ dimers on the anisotropic 
triangular
 lattice~\cite{Tamura1991, Kino1996, Nakamura2009, Kandpal,Jeschke2012}, which is 
equivalent to a square lattice model with an additional coupling along one of 
the diagonals. Many theoretical methods have been applied to the dimer based 
Hubbard model, for instance, the fluctuation-exchange 
approximation (FLEX)~\cite{Schmalian1998, Kino1998, Kondo1998, Benali2013}, the 
path-integral renormalization group~\cite{Morita2002}, cluster dynamical mean 
field theory~\cite{Parcollet2004, Tremblay2006, Hebert2015}, variational Monte 
Carlo~\cite{Liu2005, Watanabe2006,Tocchio2013,Tocchio2014} and exact 
diagonalization~\cite{Koretsune2007, Clay2008}. These studies do not agree 
entirely on all details of the phase diagram, especially whether 
superconductivity is realized in the model or not. Those studies that do show 
superconductivity nevertheless agree, with the exception of 
Ref.~\onlinecite{Benali2013}, that it is of $d_{x^2-y^2}$-type as in 
high-temperature cuprate superconductors~\cite{Scalapino2012}.

Although the triangular lattice Hubbard model has been remarkably successful in 
explaining the overall phase diagram~\cite{Lefebvre2000, Toyota2007, 
Miyagawa2004} and also some more subtle physics~\cite{EndgroupDisorder} of 
$\kappa$-(ET)$_2 X$ materials, the recent discovery of 
multiferroicity~\cite{Multiferroic} in antiferromagnetic 
$~\kappa$-(ET)$_2$Cu[N(CN)$_2$]Cl and the still unresolved problem of 
superconducting pairing-symmetry~\cite{Kuroki2006, Ardavan2012} call for 
alternative approaches. In particular, the dimer model on the anisotropic 
triangular lattice is only an approximation with {\it a priori} unclear range of 
applicability to the real lattice structure of $\kappa$-(ET)$_2 X$ charge 
transfer salts. In a seminal paper~\cite{Kuroki2002} Kuroki {\it et al.} 
investigated the superconducting pairing taking into account the realistic 
lattice structure and in fact found a phase transition between $d_{x^2-y^2}$- 
and $d_{xy}$-symmetric states when lowering the degree of dimerization. Other 
possible directions of future theoretical research beyond the dimer Hubbard 
model are outlined in Refs.~\onlinecite{Gomi2010, Hotta2010, Shinaoka2012, 
Koretsune2014}.

\begin{figure}[t]
\includegraphics[width=\linewidth]{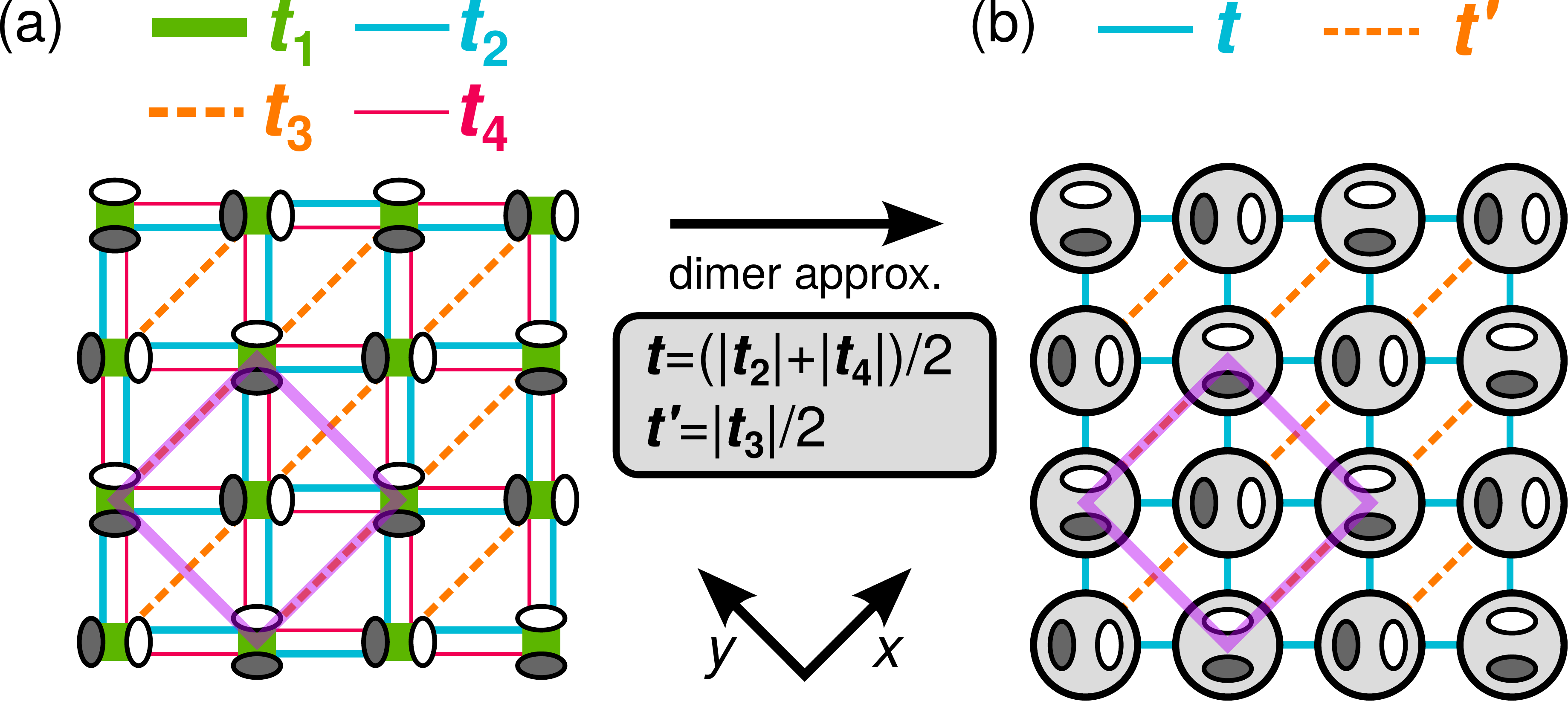}
\caption{(Color online) (a) Molecule model of the $\kappa$-(ET)$_2^+$ layer. 
Individual ET molecules are represented by white and dark grey ellipsoids. The 
four dominant hopping integrals are ($t_1$,$t_2$,$t_3$,$t_4$). Note the 
pronounced asymmetry in magnitude between $t_2$ and $t_4$, which is indicated 
here by different line thickness. (b) Dimer approximated $\kappa$-(ET)$_2^+$ 
layer. Two molecules are contracted into one dimer site indicated by a bold 
shaded circle. The intra-dimer hopping integral $t_1$ is integrated out, while 
$t_2$ and $t_4$ are averaged. Therefore, the dimer model is characterized by 
only two hopping parameters \{$t$,$t^\prime$\}=\{$(|t_2|+|t_4|)/2$,$|t_3|/2$\}. 
In both subfigures the unit cell considered in our work is indicated by a bold 
magenta colored line.}
\label{fig:lattice}
\end{figure}

In this work, building upon the idea by Kuroki {\it et al.},
 we derive a set of realistic molecule-based low-energy models
 for superconducting $\kappa$-(ET)$_2 X$ materials 
from {\it ab initio} density functional theory (DFT) calculations. After 
identifying the parameter region relevant for the real materials, we investigate 
the symmetry of the superconducting pairing in this model within a random phase 
approximation (RPA) spin-fluctuation approach. Our results show that the 
position of many materials in the phase diagram is close to a phase-transition 
line between states with extended $s+d_{x^2-y^2}$ and $d_{xy}$ pairing symmetry. 
Furthermore, we clarify that the customary dimer model not only fails in the 
limit of weak dimerization, but also when the in-plane anisotropy of hopping 
integrals becomes too large, which we find to be the case for all investigated 
materials. Finally, we simulate tunneling spectra in the superconducting state 
for selected cases and compare our findings to relevant
experiments.

\section{Methods and Models}
\subsection{Ab-initio calculations and model Hamiltonian}
\label{sec:methodsabinitio}
We use {\it ab initio} density functional theory (DFT) calculations within an 
all-electron full-potential local orbital (FPLO)\cite{FPLOmethod} basis to 
calculate the electronic bandstructure. For the exchange-correlation functional 
we employ the generalized gradient approximation 
(GGA)\cite{PerdewBurkeErnzerhof}. All calculations are converged on 
$6\times6\times6$ $k$-point grids. We use crystal structures from 
Refs.~\onlinecite{Hiramatsu, DualLayer1, DualLayer2}. In the case of 
Ref.~\onlinecite{Hiramatsu}, where crystal structures were measured for several 
temperatures, we use the data taken at 100~K.

In contrast to the customary dimer approximation, we model the 
$\kappa$-(ET)$_2^+$ layer taking into account each individual ET molecule as a 
lattice site (see Fig.~\ref{fig:lattice}). Tight-binding parameters are obtained 
from projective molecular orbital Wannier functions as implemented in 
FPLO\cite{FPLOtightbinding}. Therefore, the number of bands in the tight-binding 
model is equal to the number of ET molecules in the crystallographic unit cell. 
With the molecular Wannier function method, almost perfect representations of 
the DFT bandstructures can be obtained and ambiguities from fitting procedures 
are avoided. The latter is especially important for many-body calculations based 
on the obtained low-energy Hamiltonians.

In the following model investigation, we only keep the four largest in-plane 
hopping elements ($t_1$,$t_2$,$t_3$,$t_4$) between ET molecules [see 
Fig.~\ref{fig:lattice}(a)]. The resulting hopping structure is a generalization 
of the Shastry-Sutherland lattice~\cite{Shastry1981}, which is reached in the 
limit of $t_2 = t_4$ and $t_3 = 0$. In cases where the unit cell contains 
multiple $\kappa$-type layers, we discard all but one of the layers after the 
Wannierization procedure, because the interlayer coupling is negligible. In some 
of the investigated compounds, the crystal symmetry is lowered with respect to 
the high-symmetry orthorhombic space group {\it Pnma} of 
$~\kappa$-(ET)$_2$Cu[N(CN)$_2$]Br, which leads to a small additional splitting 
of the hoppings $t_i$ into $\tilde t_i$ and $\tilde t_i^\prime$. For simplicity, 
this particular anisotropy is discarded in our study by averaging the hopping 
integrals as $t_i = (\tilde t_i + \tilde t_i^\prime)/2$. As a result, we obtain 
the kinetic part of a four-band Hamiltonian which is 3/4-filled and of the same 
form for all materials investigated.
\begin{equation}
H_0 = \sum_{ij\sigma} t_{ij}(c^\dagger_{i \sigma} c^{\,}_{j \sigma} + h.c.)
\label{eq:noninthamiltonian}
\end{equation}

Alternatively, because ET molecules in $\kappa$-type arrangement are quite 
strongly dimerized, it is popular to approximate the $\kappa$-(ET)$_2^+$ layer by dimers 
on an anisotropic triangular lattice, integrating out the intra-dimer degrees of 
freedom. The parameters of this dimer model can be calculated directly from the 
molecule model using geometric formulas~\cite{Tamura1991}.
\begin{subequations}
\begin{align}
t & = (|t_2|+|t_4|)/2 \\
t^\prime &= |t_3|/2
\end{align}
\label{eq:geometricformulas}
\end{subequations}
By convention the dimer approximation uses the crystallographic unit cell 
containing two dimers [see Fig.~\ref{fig:lattice}(b)]. Therefore, the 
dimer-approximated Hamiltonian consists of two bands, which are half-filled.
Note that based on the geometric formulas, any anisotropy between $t_2$ and 
$t_4$ of the molecule model is discarded when going from the molecule to the 
dimer model. With few exceptions~\cite{Altmeyer2015}, the dimer approximated 
model nevertheless reproduces well the low-energy part of the original 
bandstructure. It has recently been demonstrated that improved estimates for 
dimer model parameters can be obtained by a Wannier function 
calculation~\cite{Nakamura2009, Jeschke2012, Koretsune2014}. 

\begin{figure}[t]
\includegraphics[width=\linewidth]{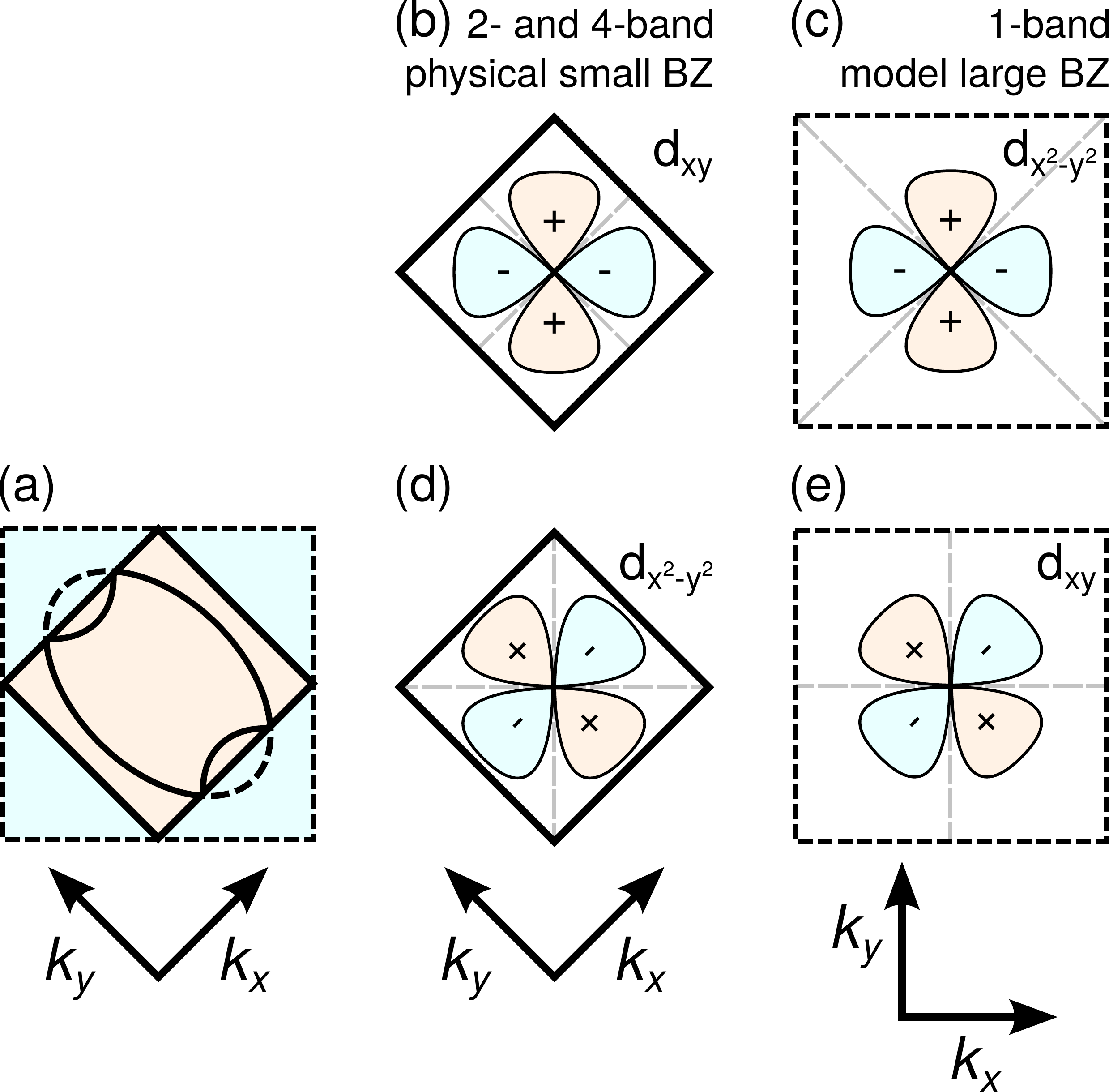}
\caption{(Color online) (a) The inner bold lines show Brillouin zone and Fermi 
surface of a generic $\kappa$-(ET)$_2 X$ material. The outer dashed lines show 
Brillouin zone and Fermi surface of the unfolded one-band dimer model.
(b) d$_{xy}$ order parameter in the physical Brillouin zone. Nodes are located 
in the $x$- and $y$-directions.
(c) d$_{x^2-y^2}$ order parameter in the unfolded Brillouin zone. Nodes are 
located along the Brillouin zone diagonals. The different designation is only 
due to a rotation of the coordinate axes by $45^\circ$.
(d) d$_{x^2-y^2}$ order parameter in the physical Brillouin zone.
(e) d$_{xy}$ order parameter in the unfolded Brillouin zone.
}
\label{fig:brillouinzones}
\end{figure}

The two-band dimer model can be unfolded to a one-band model by transforming to 
a unit cell of half the size and rotated by 45 degrees. The so-obtained model is 
directly related to the square lattice Hubbard model, but with an additional 
coupling along one of the diagonals. Results obtained in the one-band model are 
therefore rotated by 45 degrees with respect to the physical Brillouin zone of 
organic charge transfer salts, so that e.g. different $d$-wave order parameters 
exchange their designation when going from one to the other Brillouin zone (see 
Fig.~\ref{fig:brillouinzones}). Thus, the same physical order parameter which 
has $d_{xy}$-symmetry in the realistic four molecule/two dimer unit cell 
[Fig.~\ref{fig:brillouinzones}(b)] has $d_{x^2-y^2}$-symmetry in the model one 
dimer/one band unit cell [Fig.~\ref{fig:brillouinzones}(c)]. In our study, we 
always work in the physical unit cell containing two dimers 
[Fig.~\ref{fig:brillouinzones}(b) and (d)]. We refer to the small backfolded 
part of the Fermi surface close to the Brillouin zone boundary as the {\it 
elliptic} part of the Fermi surface, while we call those sheets running almost 
parallel to the $k_y$-direction {\it quasi-1D}.

An overview of unit cell and hopping paths for molecule and dimer model is shown 
in Fig.~\ref{fig:lattice}. The resulting Hamiltonians in orbital-space for all 
three cases are listed in appendix~\ref{sec:kinhamil}.

\subsection{RPA spin-fluctuation calculations}
In $\kappa$-(ET)$_2 X$ materials there is strong evidence for antiferromagnetic 
spin-fluctuations~\cite{PowellMcKenzieSpinFluctuationsNMR}. Therefore, we 
investigate the superconducting state of these materials based on a random phase 
approximation (RPA) spin-fluctuation approach in the singlet 
channel~\cite{NJPSpinfluctuationMethod,KreiselSpinFluctuations}. We have 
extended our implementation from single-site multi-orbital 
models~\cite{IntercalateRPA, K122CTRPA} to multi-site single-orbital models 
relevant for the materials discussed here. Compared to the FLEX
approximation used in Ref.~\onlinecite{Kuroki2002}, our RPA method uses only states
at the Fermi level and neglects the electronic self-energy correction. While
this approximation prevents us from making quantitative statements about the 
superconducting transition temperature $T_c$, it reduces significantly the 
numerical cost compared to FLEX, so that we can calculate the momentum structure of the
superconducting order parameter for numerous input parameter sets and 
with high angular resolution. Competing magnetically ordered or 
paramagnetic Mott insulating states are not investigated in our study. 
Furthermore, we do not investigate possible time-reversal symmetry-breaking 
superconducting states or spin-triplet pairing. 

The low-energy Hamiltonian is given by the kinetic part $H_0$, derived with the 
Wannier function method described above, and the intra-orbital Hubbard 
interaction $H_\mathrm{int}$. 
\begin{equation}
\begin{array}{rl}
H =& H_0 + H_\mathrm{int} \\ =& \sum\limits_{ij\sigma} 
t_{ij}(c^\dagger_{i \sigma} c^{\,}_{j \sigma} + h.c.)
+ \frac{{U}}{2} \sum\limits_{i\sigma} 
n_{i\sigma} n_{i{\bar \sigma}} 
\end{array}
\label{eq:hamiltonian}
\end{equation}

Here, $\sigma$ represents the spin and $n_{i\sigma} = 
c^\dagger_{i\sigma}c_{i\sigma}$. The sum over $i$ runs over all ET sites in the 
unit cell. The interaction strength $U$ is treated as a parameter. Note that 
the Coulomb repulsion on a dimer and the Coulomb repulsion on a molecule are not identical.
Especially the role of intermolecular Coulomb repulsion is currently unclear.
The investigation of interaction terms beyond on-site repulsion is left
for future studies.

We calculate the non-interacting static susceptibility $\chi^0$, where matrix 
elements $a^l_{\mu} (\vec
k)$ resulting from the diagonalization of the initial Hamiltonian
$H_0$ connect orbital and band space denoted by indices $l$ and $\mu$
respectively. The $E_\mu$ are the eigenvalues of $H_0$ and $f(E)$ is
the Fermi function. $N$ is the number of sites in the unit cell.
\begin{equation}
\begin{array}{rl}
\chi^0_{{l_1} {l_2} {l_3} {l_4}} (\vec q) = - \frac{1}{N} \sum \limits_{\vec k, 
\mu, \nu} & a_\mu^{l_4} (\vec k) a_\mu^{l_2 *} (\vec k) a_\nu^{l_1} (\vec k + 
\vec q) a_\nu^{l_3 *} (\vec k + \vec q) \\
 &\displaystyle \times \frac{f(E_\nu (\vec k + \vec q)) - f(E_\mu (\vec 
k))}{E_\nu (\vec k + \vec q) - E_\mu (\vec k)}
\end{array}
\label{eq:nonintsuscep}
\end{equation}
In our calculation both $\vec q$ and $\vec k$ run over uniform grids spanning 
the reciprocal unit cell. Temperature enters the calculation through the Fermi 
functions. 

The fraction in Eq.~\ref{eq:nonintsuscep} becomes problematic in numerical 
calculations, when the band energies $E_\nu$ and $E_\mu$ become degenerate. 
However, the expression can be rectified using l'Hospital's rule, which we use 
in practice when the magnitude of the denominator falls below a certain 
threshold (e.g. $10^{-7}~\mathrm{eV}$). Here, $\beta$ denotes the inverse 
temperature $\beta= (k_B T)^{-1}$.
\begin{equation}
\lim \limits_{E_\nu \to E_\mu} \frac{f(E_\nu (\vec k + \vec q)) - f(E_\mu (\vec 
k))}{E_\nu (\vec k + \vec q) - E_\mu (\vec k)} = - \beta \frac{e^{\beta  
E_\nu}}{(e^{\beta E_\nu} + 1)^2}
\label{eq:nestingfunctionlHosp}
\end{equation}

The static spin- and orbital-susceptibilities ($\chi^{s,\mathrm{RPA}}$ and 
$\chi^{c,\mathrm{RPA}}$) are constructed in an RPA framework. Since the 
interaction term defined in Eq.~\ref{eq:hamiltonian} is local and we have only 
one orbital per lattice site, we can restrict the calculation to the diagonal 
elements of the susceptibility and use scalar equations for the RPA-enhanced 
susceptibilities.
\begin{subequations}
\begin{align}
\chi^{s,\mathrm{RPA}}_L ({\vec q}) =& \frac{\chi^0_L (\vec q)}{1 - {U} \chi^0_L 
(\vec q)} \\
\chi^{c,\mathrm{RPA}}_L ({\vec q}) =& \frac{\chi^0_L (\vec q)}{1 + {U} \chi^0_L 
(\vec q)}
\end{align}
\label{eq:rpasuscep}
\end{subequations}
Here, $\chi_L$ with $L=\{llll\}$ denotes the diagonal element of the 
susceptibility tensor associated with an ET site indexed by $l$. Note that this 
formulation allows us to treat multiple inequivalent ET sites in the unit cell, 
keeping the individual ${\vec q}$-dependence of their associated 
susceptibilities. Therefore, the symmetry of the susceptibility follows the 
symmetry of the ET layer in the crystallographic unit cell, which is important 
for checking the simplified four-parameter model against {\it ab initio} 
Hamiltonians, which can have monoclinic, as e.g. in 
$\kappa$-(ET)$_2$Cu(NCS)$_2$, or even triclinic symmetry, as in 
$\kappa$-$\alpha_1^\prime$-(ET)$_2$Ag(CF$_3$)$_4$(TCE).

The total spin susceptibility is given by the sum over all site-resolved 
contributions:
\begin{equation}
\chi^s ({\vec q}) = \frac{1}{2} \sum\limits_L \chi^{s,\mathrm{RPA}}_L ({\vec q})
\label{eq:obssuscep}
\end{equation}

The pairing vertex in orbital space for the spin-singlet channel can be 
calculated
using the fluctuation exchange
approximation~\cite{BickersScalapinoWhitePairingVertex,
ScalapinoLohHirsch}:
\begin{equation}
\begin{aligned}
\Gamma_{{l_1} {l_2} {l_3} {l_4}} (\vec k, \vec k^\prime) = &\left[\frac{3}{2} 
{U} \chi^{s,\mathrm{RPA}} (\vec k - \vec k^\prime) {U}\right.\\
- &\left.\frac{1}{2} {U} \chi^{c,\mathrm{RPA}} ({\vec k} - {\vec k^\prime}) {U} 
+ {U} \right]_{{l_1} {l_2} {l_3} {l_4}}
\end{aligned}
\label{eq:pairingvertexorbitalspace}
\end{equation}
In the pairing vertex, the momenta $\vec k$ and $\vec k^\prime$ are restricted to the 
Fermi surface. As vectors $\vec k - \vec k^\prime$ do not necessarily lie on the 
grid used in the calculation of the susceptibility $\chi^0 (\vec q)$, we 
interpolate the grid data linearly.

The pairing vertex in orbital space is transformed into band space using the 
matrix elements $a^l_{\mu} (\vec k)$:
\begin{equation}
\begin{aligned}
\Gamma_{\mu \nu} (\vec k, \vec k^\prime) = \mathrm{Re} \sum\limits_{l_1 l_2 l_3 
l_4} &a_\mu^{l_1,*} (\vec k) a_\mu^{l_4,*} (-\vec k)  \left[\Gamma_{{l_1} {l_2} 
{l_3} {l_4}} (\vec k, \vec k^\prime)\right]\\ \times & a_\nu^{l_2} (\vec 
k^\prime) a_\nu^{l_3} (-\vec k^\prime)
\end{aligned}
\label{eq:pairingvertexbandspace}
\end{equation} 

Finally, we solve the linearized gap equation by performing an 
eigendecomposition on the kernel and obtain the dimensionless pairing strength 
$\lambda_i$ and the symmetry function $g_i (\vec k)$.
\begin{equation}
- \sum \limits_\nu \oint_{C_\nu} \frac{dk^\prime_\parallel}{2\pi} \frac{1}{2\pi 
\, v_F (\vec k^\prime)} \left[ \Gamma_{\mu\nu} (\vec k, \vec k^\prime) \right]  
g_i (\vec k^\prime) = \lambda_i g_i (\vec k)
\label{eq:gapequation}
\end{equation}
The integration runs over the discretized Fermi surface and $v_F (\vec k)$ is 
the magnitude of the Fermi velocity.

For the computations presented in this paper, we evaluated the susceptibility 
$\chi^0 (\vec q)$ using $50 \times 50$ point grids for $\vec q$ and the 
integrated-out variable $\vec k$ (see Eq.~\ref{eq:nonintsuscep}). The inverse 
temperature in the susceptibility calculation is fixed to $\beta = 160/t_1$ for 
the molecule model and $\beta = 60/t$ for the dimer model. These values result in about 
the same effective temperature. The Fermi surface is 
determined by inverting linear interpolants for the band energies on a fine 
grid. For the models considered here about $250$ points on the Fermi surface are 
sufficient. The Hubbard repulsion parameter $U$ is chosen in all calculations so 
that the leading eigenvalue in Eq.~\ref{eq:gapequation} is $\lambda = 0.99\pm0.001$.
For most combinations of input parameters this leads to a clear separation of the
leading and the first subleading eigenvalue. The pairing symmetries
corresponding to the leading and sub-leading eigenvalues do not change as 
a function of $U$.

\begin{table*}[t]
  \caption{Values of the molecule model parameters
  ($t_1$,$t_2$,$t_3$,$t_4$), also commonly
  denoted as ($b_1$,$p$,$b_2$,$q$), for selected superconducting 
$\kappa$-(ET)$_2 X$ materials. All values are given in meV. Hopping integrals 
for the (ET)$_2$Ag(CF$_3$)$_4$(TCE) family are averages of the
  parameters given in Ref. \onlinecite{Altmeyer2015}, where we used the same 
method and settings to calculate the parameters as in the present study. Crystal 
structures for these materials were taken from Refs.~\onlinecite{DualLayer1, 
DualLayer2}. All other crystal structures are taken from 
Ref.~\onlinecite{Hiramatsu}. The values for $T_c$ are taken from 
Refs.~\onlinecite{Hiramatsu, Kato1987, Mori1990, Kini1990}. In 
Ref.~\onlinecite{EndgroupDisorder} models of the same form for 
non-superconducting $\kappa$-(ET)$_2 X$ compounds are listed.}
\label{tab:materialparameters}
\begin{ruledtabular}
\begin{tabular}{rlrrrrr|rrr|r}
$i$ &material & $T_c$ in K & $t_1$ & $t_2$ & $t_3$ & $t_4$ & $t_2/t_1$ & 
$t_3/t_1$ & $t_4/t_1$ & $t_4/t_2$ \\
\hline
 1 & $\kappa$-(ET)$_2$Ag(CF$_3$)$_4$(TCE)                     &  2.6 & 168 & 
102 & 60.8 & 33.4 & 0.610 & 0.362 & 0.199 & 0.362 \\
 2 & $\kappa$-(ET)$_2$I$_3$                                   &  3.6 & 180 & 119 
  & 52.2 & 31.7 & 0.661 & 0.289 & 0.176 & 0.266 \\
 3 & $\kappa$-(ET)$_2$Ag(CN)$_2 \cdot $H$_2$O                 &  5.0 & 185 & 104 
  & 60.4 & 23.6 & 0.567 & 0.326 & 0.173 & 0.305 \\
 4 & $\kappa$-$\alpha_1^\prime$-(ET)$_2$Ag(CF$_3$)$_4$(TCE)   &  9.5 & 166 &  
97.6 & 65.8 & 35.3 & 0.588 & 0.396 & 0.213 & 0.362 \\
 5 & $\kappa$-(ET)$_2$Cu(NCS)$_2$                             & 10.4 & 190 & 102 
  & 82.4 & 17.5 & 0.538 & 0.387 & 0.092 & 0.171 \\
 6 & $\kappa$-$\alpha_2^\prime$-(ET)$_2$Ag(CF$_3$)$_4$(TCE)   & 11.1 & 165 &  
98.4 & 66.7 & 36.3 & 0.596 & 0.404 & 0.220 & 0.369 \\
 7 & $\kappa$-(ET)$_2$Cu[N(CN)$_2$](CN)                       & 11.2 & 175 & 100 
  & 78.5 & 17.3 & 0.574 & 0.344 & 0.099 & 0.172 \\
 8 & $\kappa$-(ET)$_2$Cu[N(CN)$_2$]Br                         & 11.6 & 177 &  
95.6 & 60.0 & 36.2 & 0.541 & 0.339 & 0.205 & 0.379 \\
\end{tabular}
\end{ruledtabular}
\end{table*}

\subsection{Simulation of tunneling spectra in the superconducting state}
The central quantity measured in the scanning tunneling spectroscopy (STS) 
experiments on superconductors is the local density of states (DOS) in 
the superconducting phase. Here we start from the standard Bardeen-Cooper-Schrieffer 
(BCS) theory for isotropic s-wave superconductors. A simple approximate 
extension allows us to treat realistic Fermi surfaces and unconventional pairing 
symmetries derived from the {\it ab initio} calculations combined with RPA 
spin-fluctuation theory as presented above.

To derive an approximation for the DOS of a superconductor, we start with the 
Hamiltonian for Cooper pairs with vanishing total momentum~\cite{Bardeen1957}.
\begin{equation}
H = \sum_{k,\sigma} \epsilon_{k\sigma} c_{k\sigma}^\dagger c_{k\sigma} + 
\sum_{k,k'} U(k,k') c_{k\uparrow}^\dagger c_{-k\downarrow}^\dagger 
c_{-k'\downarrow} c_{k'\uparrow}
\label{eq:bcshamil}
\end{equation}
The interaction can be treated in mean field theory ($\delta(c^\dagger 
c^\dagger)=c^\dagger c^\dagger-\langle c^\dagger c^\dagger \rangle$), where 
terms quadratic in $\delta$ are neglected. The resulting Hamiltonian can be 
diagonalized using the Bogoliubov-Valatin transformation which introduces 
quasiparticle creation and annihilation operators $\gamma_{k\sigma}^\dagger$ and 
$\gamma_{k\sigma}$. The quasiparticle excitation energies are given as 
$E_k=\sqrt{\epsilon_k^2+\vert \Delta_k \vert^2}$, where $\Delta(k)=\sum_{k'} 
U(k,k')\langle c_{-k'\downarrow} c_{k'\uparrow}\rangle$.

The BCS Hamiltonian can be rewritten in terms of the quasiparticle creation and 
annihilation operators:
\begin{equation}
\begin{array}{rl}
H_{\textnormal{BCS}} =& \sum_{k,\sigma} E_k  \gamma_{k\sigma}^\dagger 
\gamma_{k\sigma}+ \sum_{k} \epsilon_{k} \\[4pt]
&- \sum_{k,k'} U(k,k')\langle c_{k\uparrow}^\dagger 
c_{-k\downarrow}^\dagger\rangle \langle c_{-k'\downarrow} c_{k'\uparrow} \rangle
\end{array}
\label{eq:bcshamilqp}
\end{equation}
The excitation spectrum of the quasiparticles $E_k$ is gapped and defined only 
for positive energies.
The density of states of quasiparticles in an isotropic s-wave superconductor 
can be calculated from the normal state density of states $\rho (\epsilon )$ and 
the constant superconducting gap $\Delta_k=\Delta$:
\begin{equation}
\begin{array}{rl}
\rho_{\textnormal{qp}}(E)=&\frac{1}{N} \sum_k \delta(E-E_k)\\[4pt]
=&\int d\epsilon \, \rho_0(\epsilon) \frac{\sqrt{\epsilon^2+\vert 
\Delta\vert^2}}{\epsilon} \delta(\epsilon -\sqrt{E^2-\vert 
\Delta\vert^2})\\[4pt]
=&\begin{cases} \rho_0(\sqrt{E^2-\vert \Delta\vert^2}) \frac{E}{\sqrt{E^2-\vert 
\Delta\vert^2}} & E > \vert \Delta\vert\\[4pt]
0 & E < \vert \Delta\vert
\end{cases}
\end{array}
\label{eq:dosqpswave}
\end{equation}

The previous derivation assumed an isotropic gap and an energy dispersion of 
free electrons to identify the normal state DOS $\rho_0$.
For realistic electronic structure and anisotropic gap $\Delta_k$ this 
factorization of contributions is not easily possible due to the non-trivial 
momentum dependence of both functions:
\begin{equation}
 \begin{array}{rl}
\rho_{\textnormal{qp}}(E) =&\int d\epsilon   \frac{1}{N} \sum_k \delta( \epsilon-\epsilon_k) 
\delta(\vert E\vert -\sqrt{\epsilon^2+\vert \Delta_k\vert^2})\\[4pt]
\neq&\int d\epsilon \, \rho_N (\epsilon) \, \delta(\vert E\vert 
-\sqrt{\epsilon^2+\vert \Delta_k\vert^2})
\end{array}
\label{eq:dosqpapprox}
\end{equation}
However, in a widely used ansatz~\cite{Tanaka1995,Hasegawa1996} the electrons with effective mass $m^\ast$ are considered 
to be free, i.e. the Fermi surface is approximated
by a concentric circle, and the gap only 
depends on the angle $\theta$.
 \begin{equation}
\rho_{\textnormal{qp}}(E) \approx  \frac{1}{(2 \pi)^2} m^\ast \textnormal{Re} \int d\theta  
\frac{\vert E \vert}{\sqrt{E^2-\vert \Delta(\theta)\vert^2}}
\label{eq:dosqpapproxangleintegration}
\end{equation}

We introduce in this expression a finite quasiparticle lifetime~\cite{Dynes1978} 
by adding an imaginary part $\Gamma$ to the quasiparticle excitation energies. 
This allows us to carry out calculations with finite angular resolution and 
facilitates comparison to experiment. Furthermore, we improve upon the circular 
integration by replacing it with a summation over the discretized realistic 
Fermi surface and drop the irrelevant prefactors to obtain the final expression 
for the quasiparticle DOS in our study:
\begin{equation}
\rho_{\textnormal{qp}}(E) \propto \sum\limits_{\vec k} \,\textnormal{Re}\frac{ \vert 
E+i\Gamma \vert}{\sqrt{(E+i\Gamma)^2-\Delta(\vec k)^2 }}
\label{eq:dosqpfinal}
\end{equation}
In this form, the connection to the {\it ab initio} and RPA spin-fluctuation 
calculations is easily obtained: the vectors $\vec k$ in Eq.~\ref{eq:dosqpfinal} 
all lie on the Fermi surface determined from the {\it ab initio} derived 
tight-binding model and the gap $\Delta (\vec k)$ on the Fermi surface can be 
substituted by the symmetry function $g_i (\vec k)$ extracted from RPA 
(Eq.~\ref{eq:gapequation}). Note that the overall energy scale of the 
superconducting gap is not included in $g_i (\vec k)$ because our formalism 
neglects the electronic self-energy and lacks a self-consistency condition. We 
checked that our approximation agrees well with a direct calculation of the 
quasiparticle spectrum based on Eq.~\ref{eq:bcshamilqp}, which is numerically 
more costly.

The quasiparticle DOS $\rho_{\textnormal{qp}} (E)$ corresponds to the local density of states 
(LDOS) measured in STS experiments. For a direct comparison, thermal smearing 
and additional effects such as electronic disorder might have to be taken into 
account~\cite{Diehl2014, Diehl2016, EndgroupDisorder}.

\section{Results and Discussion}
\subsection{Ab-initio  calculations}
Using {\it ab initio} density functional theory calculations and subsequent 
Wannier downfolding we determine the parameter sets ($t_1$,$t_2$,$t_3$,$t_4$) 
corresponding to superconducting $\kappa$-(ET)$_2 X$ materials with anions {\it 
X}$\,\in\,$\{Ag(CF$_3$)$_4$(TCE), I$_3$, Ag(CN)$_2 \cdot $H$_2$O, Cu(NCS)$_2$, 
Cu[N(CN)$_2$](CN), Cu[N(CN)$_2$]Br\}, as well as polymorphs 
$\kappa$-$\alpha_1^\prime$-(ET)$_2$Ag(CF$_3$)$_4$(TCE) and 
$\kappa$-$\alpha_2^\prime$-(ET)$_2$Ag(CF$_3$)$_4$(TCE), which also contain 
charge-ordered insulating $\alpha^\prime$-type layers. The calculated parameters 
are listed in Table~\ref{tab:materialparameters}. In the case of 
(ET)$_2$Ag(CF$_3$)$_4$(TCE) polymorphs (TCE abbreviates 1,1,2-trichloroethane) 
we rely on a previous {\it ab initio} calculation with identical 
setup~\cite{Altmeyer2015}. The small asymmetry of hoppings due to the lowered 
symmetry in these materials is averaged out to obtain a four-parameter model. 
For the original models, see Ref.~\onlinecite{Altmeyer2015}.

We observe that all materials fall into a narrow region of parameters: $t_1 
\in$~[165,190]~meV, $t_2 \in$~[95.6,119]~meV, $t_3 \in$~[52.2,82.4]~meV and $t_4 
\in$~[17.3,36.3]~meV. Normalizing $t_2$, $t_3$ and $t_4$ with respect to $t_1$, 
this means all materials lie in the range $t_2/t_1~\in$~[0.538,0.661], 
$t_3/t_1~\in$~[0.289,0.404] and $t_4/t_1~\in$~[0.099,0.220]. Note the pronounced 
anisotropy between $t_2$ and $t_4$. These intervals of $t_2/t_1$, $t_3 / t_1$ 
and $t_4 / t_1$ obtained from the {\it ab initio} calculations determine the 
parameter ranges for our following model investigation.

We sorted the materials according to their superconducting transition 
temperature $T_c$, but we found no correlation of $T_c$ with either $t_1$, 
$t_2$, $t_3$ or $t_4$. The ratios $t_2/t_1$, $t_3/t_1$ or $t_4/t_1$ are also not 
obviously connected to $T_c$.

\begin{figure*}[t]
\includegraphics[width=\linewidth]{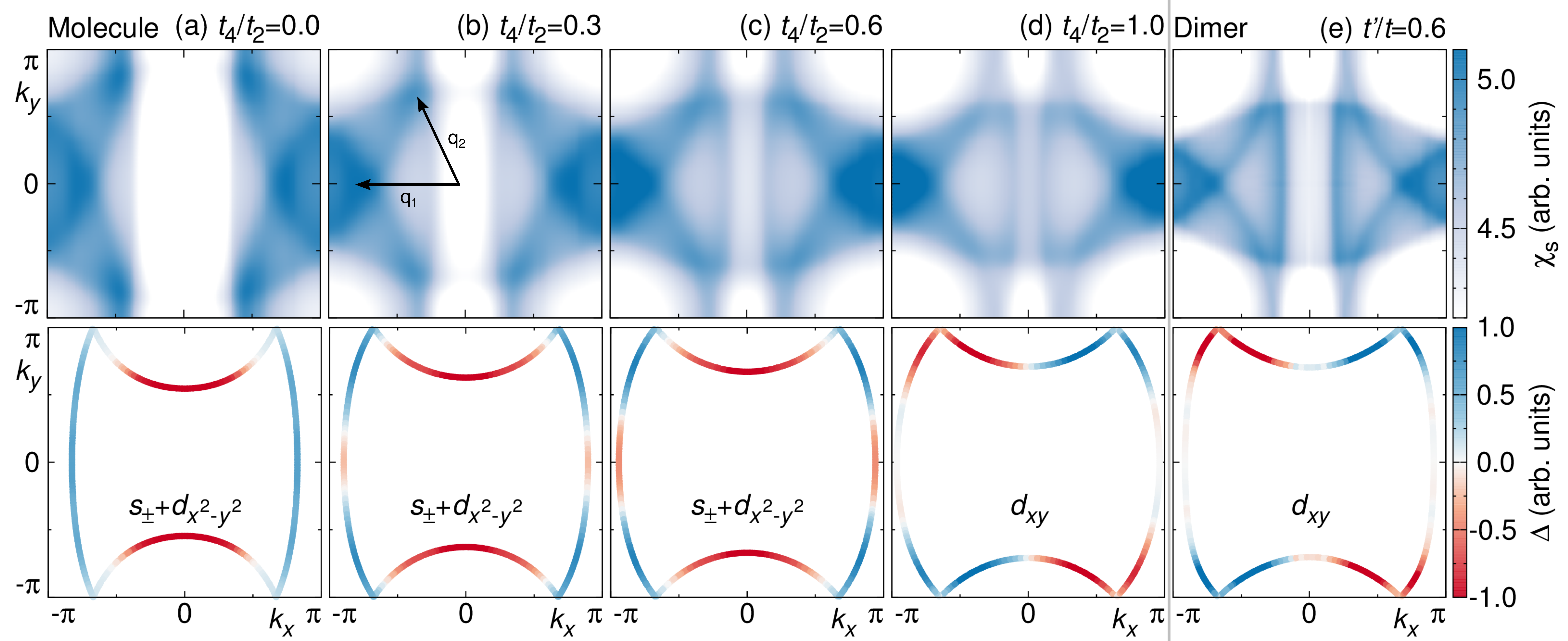}
\caption{(Color online) Comparison of molecule models (a)-(d) with different 
ratios of $t_4/t_2$, which all correspond to the same dimer model (e) with 
$t^\prime/t = 0.6$. The top panel shows the spin susceptibilities, where arrows $\vec q_1$ 
and $\vec q_2$ indicate the main features, while the 
bottom panel shows the leading eigenfunction of the superconducting gap equation 
on the Fermi surface. In the molecule models $t_3/t_1 = 0.333$ is fixed, while 
the ratio of $t_4/t_2$ is varied under the condition $t_3/(t_2+t_4) = t^\prime / 
t = 0.6$.}
\label{fig:susceptibilitycomparison}
\end{figure*}

\subsection{Pairing symmetry in the dimer model}
First, we apply the RPA spin-fluctuation formalism to the dimer model in the 
range $t^\prime / t \in$~[0,1].
 We evaluate the superconducting order parameter 
in fine steps of $t^\prime / t$ and compare the leading eigenfunctions. In all 
cases we find that a $d_{xy}$-state is the leading pairing symmetry.

Relating the dimer model back to the one-band model explained in the methods 
section, the $d_{xy}$-state we find is identical to the $d_{x^2-y^2}$-state of 
the square lattice Hubbard model after unfolding the Brillouin zone (see 
Fig.~\ref{fig:brillouinzones}). Typical superconducting $\kappa$-(ET)$_2 X$ 
materials lie in the region $t^\prime / t \lesssim 0.65$~\cite{Kandpal, 
Nakamura2009, EndgroupDisorder}, where a $d_{xy}$-solution is to be expected, as 
the dimer model is basically a square lattice of hoppings $t$, perturbed by the 
additional diagonal coupling $t^\prime$. The diagonal coupling $t^\prime$ breaks 
the $C_4$-symmetry of the underlying square lattice and gives the Fermi surface 
its elliptic shape, but the dominant terms in the Hamiltonian remain square 
lattice-like. For a full account of possible pairing symmetries in the one-band 
Hubbard model on the square lattice, see Ref.~\onlinecite{Roemer2015}.

An early theoretical study of the antiferromagnetic phase of $\kappa$-type 
materials concluded that the two molecules within a dimer carry the same spin 
and that the spins are flipped between neighboring dimers~\cite{Kino1996}, 
giving rise to $(\pi, \pi)$ magnetic order as in the parent compounds of 
high-temperature cuprate superconductors~\cite{Scalapino2012}. This result is 
consistent with our observation that a dimerized model gives a $d_{xy}$ order 
parameter in the physical Brillouin zone, which becomes a $d_{x^2-y^2}$-symmetry 
in the unfolded zone of the one-band model (see Fig.~\ref{fig:brillouinzones}), 
again emphasizing the deep connection between cuprates and quasi-two-dimensional 
organic superconductors.

We would also like to note that the authors of Ref.~\onlinecite{Oka2015}, 
referring to the physical Brillouin zone, invoked a $d_{xy}$ superconducting 
symmetry close to insulating patches and a $d_{x^2-y^2}$ state in the bulk to 
explain the findings of their STS study on deuterated 
$\kappa$-(ET)$_2$Cu[N(CN)$_2$]Br. As the antiferromagnetic insulating state is 
dimerized (see Ref.~\onlinecite{Kino1996}), the dimer approximation naturally 
applies and gives a $d_{xy}$ order parameter in accordance with the experimental 
observation. What remains to be answered in an approach beyond the dimer model,
as presented in the next sections, 
is why the superconducting order parameter of the bulk is $d_{x^2-y^2}$.

\subsection{Pairing symmetry in the molecule model}
\begin{figure}[t]
\includegraphics[width=0.75\linewidth]{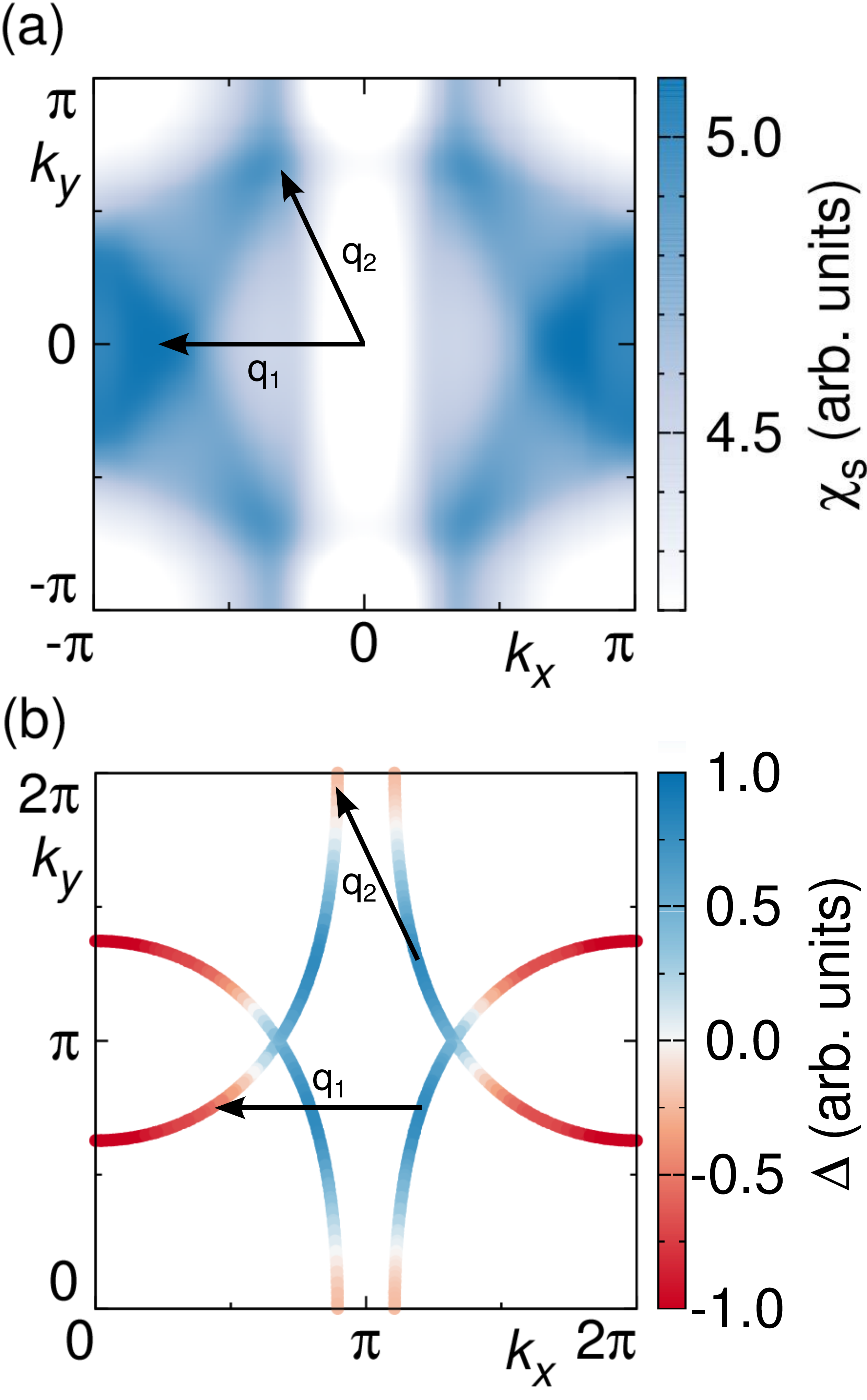}
\caption{(Color online) (a) Spin susceptibility of the molecule model at $t_2/ 
t_1 = 0.417$, $t_3 / t_1 = 0.333$ and $t_4 / t_1 = 0.139$ with arrows $\vec q_1$ 
and $\vec q_2$ indicating the main features. The data shown are the same as in 
Fig.~\ref{fig:susceptibilitycomparison}(b). (b) Superconducting gap function on 
the Fermi surface for the same parameter values as in (a). Vectors $\vec q_1$ 
and $\vec q_2$ are the same as in (a) and connect parts of the Fermi surface 
with different signs of the gap. 
Note that the plot range of the Brillouin zone is shifted by a vector $(\pi, 
\pi)$ compared to Fig.~\ref{fig:susceptibilitycomparison}.}
\label{fig:susceptibilityvectors}
\end{figure}

The obvious step for going beyond the dimer model is to use the original crystal 
lattice, i.e. the molecule model explained above (see Fig.~\ref{fig:lattice}).
In order to compare the results of the molecule model to those of the
dimer model we do the following consideration: 
via the geometric Eq.~\ref{eq:geometricformulas} the four-parameter molecule 
model is mapped onto a two-parameter dimer model. With this procedure
we are left with two adjustable
parameters in the molecule model whose variation discloses important 
features of the systems,  not
captured in the  resulting dimer model which remains unchanged. 
 These adjustable parameters are: the 
degree of dimerization $t_1/\text{max}(t_2, t_3, t_4)$ and the in-plane 
anisotropy $t_4 / t_2$. 

The degree of dimerization obviously decides whether the dimer approximation 
applies to a material or not. Its influence on the superconducting pairing was 
quantified by Kuroki {\it et al.}, who found a transition to a $d_{x^2-y^2}$ 
state in the physical Brillouin zone at low dimerization (see 
Ref.~\onlinecite{Kuroki2002}).

Less obvious is how this $d_{x^2-y^2}$ state emerges from the underlying hopping 
structure and how important the anisotropy between $t_2$ and $t_4$ is for the 
pairing state. To investigate these issues, we construct a series of molecule 
models with fixed value of $t_3 / t_1 = 0.333$ and vary the ratio $t_4 / t_2$ in 
the range [0,1] (for the values realized in real materials see 
Table~\ref{tab:materialparameters}). We fix the sum of $t_2$ and $t_4$ so that 
the molecule models correspond to the same dimer model $t_3/(t_2+t_4) = t^\prime 
/ t = 0.6$. The maximum value of $t_2$ is therefore $t_2^\text{max} = 0.556$ and 
its minumum value is $t_2^\text{min} = 0.278$. As $t_4$ is increased, the 
in-plane anisotropy decreases and the dimerization defined as 
$t_1/\text{max}(t_2, t_3, t_4)$ increases.

In Fig.~\ref{fig:susceptibilitycomparison} we show the spin susceptibilities and 
leading pairing symmetries in the molecule model as a function of $t_4 / t_2$ 
compared to the associated dimer model. In the isotropic limit $t_4 / t_2 = 1$ 
we find a $d_{xy}$-symmetric state, similar to the one found in the dimer 
model [compare  Fig.~\ref{fig:susceptibilitycomparison}(d) and  
Fig.~\ref{fig:susceptibilitycomparison}(e)]. Upon lowering  $t_4 / t_2$ the 
maxima of the superconducting gap shift toward the position where the nodes in a 
$d_{xy}$-symmetric state are and additional nodes appear on the quasi-1D part of 
the Fermi surface close to $(\pm \pi, 0)$. As this shift is equivalent to a 
rotation by 45 degrees, the state with eight nodes can be expected to have 
significant $d_{x^2-y^2}$ contribution. In the limit of $t_4 \ll t_2$ the 
additional set of nodes on the quasi-1D part of the Fermi surface vanishes [see  
Fig.~\ref{fig:susceptibilitycomparison}(a)]. The remaining four nodes are 
situated close to the Brillouin zone boundary, where the smaller elliptic part 
of the Fermi surface is folded back.  The details of the pairing symmetry are 
discussed further below. In what follows we investigate the origin of the gap 
maxima shifts.

As an example we show in Fig.~\ref{fig:susceptibilityvectors} how extrema of the 
gap magnitude with opposite sign appear where parts of the Fermi surface can be 
connected by a wave-vector $\vec q$ that shows a peak in the spin 
susceptibility. Note that in Fig.~\ref{fig:susceptibilityvectors} the Brillouin 
zone is shifted by a vector $(\pi, \pi)$, because the relevant vectors $\vec q$ connect 
pieces of the Fermi surface across the boundaries of the Brillouin zone 
used in Fig.~\ref{fig:susceptibilitycomparison}.

Now we come back to the discussion of the results presented in 
Fig.~\ref{fig:susceptibilitycomparison}. At $t_4/t_2 = 0$ peaks appear at $\vec 
q_1 \approx (\pm 0.7 \pi, 0)$ and $\vec q_2 \approx (\pm \pi/2, \pm \pi)$, the 
dominant contribution to the spin susceptibility being the peak at $\vec q_2$. 
As $t_4 / t_2$ is increased, the position of $\vec q_1$ remains about the same, 
while $\vec q_2$ shifts towards $(\pm \pi/4, \pm \pi/2)$ and decreases in 
intensity. At $t_4 / t_2 = 1$, the peak at $\vec q_1$ becomes the dominant 
contribution to the spin susceptibility. As we do not work in the limit of 
infinite dimerization, even the case $t_4 / t_2 = 1$ does not reproduce the 
dimer model spin susceptibility exactly. The similarities are however apparent. 

These peak shifts in the spin susceptibility are reflected in the pairing 
symmetry: the gap maxima of different sign in the $d_{xy}$-symmetry are 
separated by a wave-vector $\vec q_1$, while $\vec q_2$ is responsible for the 
sign change between the upper and lower half of the elliptic Fermi surface. 
Furthermore, $\vec q_2$ enforces an enlarged nodal region close to $(\pm \pi, 
0)$, since it would otherwise connect parts of the Fermi surface with the same 
sign of the gap. In the intermediate region of $t_4 / t_2$, $\vec q_2$ connects 
the 1D parts of the Fermi surface, where it induces an additional set of nodes. 
The large gap on the elliptic part of the Fermi surface is connected to the 
1D sheets by $\vec q_1$. The shift of the vertical lines in the susceptibility, 
which widen towards $k_x \approx \pm \pi/2$, merely reflect the changing shape 
of the Fermi surface. For $t_4 \ll t_2$ the additional set of nodes on the 
1D sheets vanishes, because they are no longer connected by $\vec q_2$, which 
now instead points from 1D sheet to the elliptic parts just like $q_1$. This 
consideration shows that the pairing-symmetry transition in the molecule model 
is driven by a peculiar competition between $\vec q_1$ and $\vec q_2$ nesting 
vectors.

\begin{figure}[tb]
\includegraphics[width=0.8\linewidth]{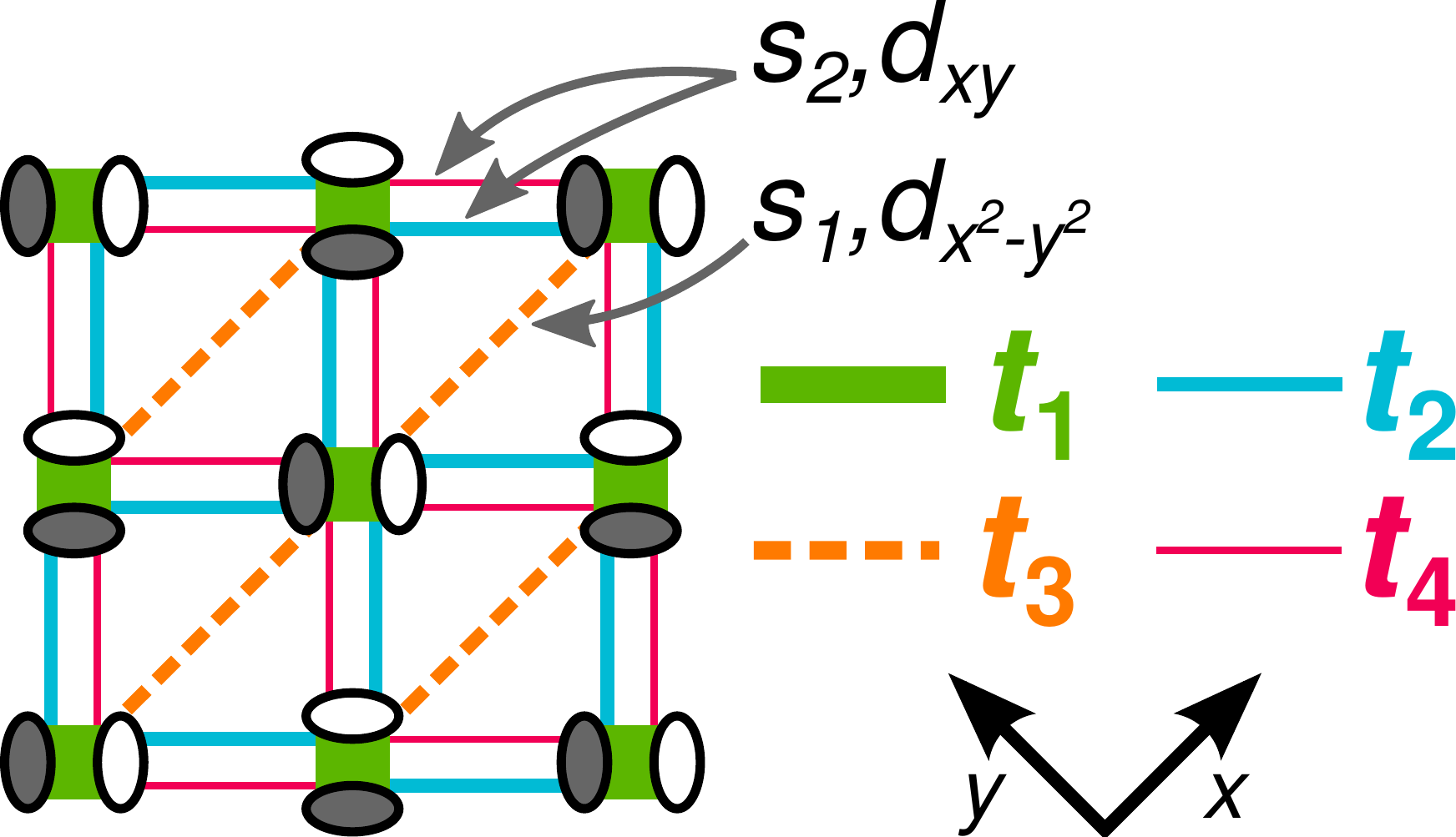}
\caption{(Color online) Hopping structure in the molecule model for the 
$\kappa$-(ET)$_2^+$ layer. The dark grey arrows indicate the connection 
between hopping parameters and the symmetry functions appearing in the 
solution of the superconducting gap equation.}
\label{fig:pairingcartoon}
\end{figure}

Now we connect the structure of the susceptibility and the superconducting 
pairing to the underlying lattice model. The feature at $\vec q_1$ is obviously 
connected to the $t_3$ hopping parameter, since it is the only hopping 
exclusively in $x$-direction (compare Fig.~\ref{fig:lattice}). All other 
$t$-parameters can only be responsible for a four-peak structure, as they occur 
pointing along both diagonals of the physical unit cell. The influence of the 
competition between $t_2$, $t_3$ and $t_4$ on the feature at $\vec q_2$ is 
however hard to quantify directly. Therefore, we decompose the superconducting 
order parameter in terms of extended $s$- and $d$-wave basis functions $f_i$ 
appropriate for a square lattice geometry. For each of the $d$-wave basis 
functions, we also take into account the associated extended $s$-wave function, 
because we expect that a significant extended $s$-wave component could mix with 
the $d$-wave states to accomodate the orthorhombicity of the model:
\begin{subequations}
\begin{align}
f_{s_1} (\vec k)&=\text{cos} k_x + \text{cos} k_y \\
f_{d_{x^2-y^2}} (\vec k)&=\text{cos} k_x - \text{cos} k_y \\
f_{s_2} (\vec k)&=\text{cos} k_x \cdot \text{cos} k_y \\
f_{d_{xy}} (\vec k)&=\text{sin} k_x \cdot \text{sin} k_y 
\end{align}
\label{eq:symmfunctions}
\end{subequations}
Rotated into the Brillouin zone of $\kappa$-type materials, gap functions 
$f_{d_{xy}}$ and $f_{s_2}$ are to be expected from antiferromagnetic exchange 
along square-like bonds ($t_2$, $t_4$), while $f_{s_1}$ and $f_{d_{x^2-y^2}}$ 
correspond to exchange paths along diagonal bonds ($t_3$), see 
Fig.~\ref{fig:pairingcartoon}.

We fit the pairing symmetries calculated from RPA to a linear combination of 
the previously defined pairing symmetries and determine their relative 
contributions $c_i$.
\begin{equation}
\tilde g  (\vec k) = c_{s_1} f_{s_1} + c_{d_{x^2-y^2}} f_{d_{x^2-y^2}} + c_{s_2} f_{s_2} + c_{d_{xy}} 
f_{d_{xy}}
\label{eq:modelgap}
\end{equation}
For the $d_{xy}$-state we find $c_{d_{xy}} = 1$ and all other contributions 
zero, i.e. except for the not well reproduced extended nodal region close to 
$(\pm \pi, 0)$ the dimer model and the molecule model at $t_4 / t_2 \lesssim 1$ 
are dominated by the square-lattice physics of $t$ and $t_2$, $t_4$ 
respectively. For the $d_{x^2-y^2}$-like solution at $t_4 \ll t_2$ we find 
negligible contributions from $f_{s_1}$ and $f_{d_{xy}}$, dominant $f_{s_2}$ and 
sub-dominant $f_{d_{x^2-y^2}}$. For increasing $t_4 / t_2$ the ratio of 
coefficients $c_{d_{x^2-y^2}} / c_{s_2}$ decreases, i.e. the square-lattice 
physics becomes dominant when the asymmetry between $t_2$ and $t_4$ is removed. 
Using the symmetry functions $f_i$, all details of the superconducting gap in 
the $d_{x^2-y^2}$-like state including the additional nodes can be reproduced by 
Eq.~\ref{eq:modelgap}.

Our findings provide a clear picture of the pairing competition in the molecule 
model: in the realistic region of parameters, where the dimerization measured by 
$t_1/\text{max}(t_2, t_3, t_4)$ and the anisotropy of $t_2$ and $t_4$ are 
finite, the competition of square-like ($t_2$, $t_4$) and diagonal ($t_3$) 
hopping realizes a unique linear combination of functions $f_{d_{x^2-y^2}}$ and 
$f_{s_2}$ as the leading pairing symmetry. We refer to this linear combination 
as $s_\pm + d_{x^2-y^2}$, or extended $s+d_{x^2-y^2}$. The $s$-wave contribution 
is equivalent to the $s_\pm$ pairing state believed to be realized in iron-based 
superconductors (see e.g. Ref.~\onlinecite{Hosono2015}) and has been overlooked 
entirely in the literature on quasi-two-dimensional organic charge transfer 
salts. When the lattice becomes more square-like $(t_4 \lesssim t_2)$, i.e. the 
molecule model approaches the dimer limit, the $d_{xy}$-symmetry known from the 
dimer model takes over. In other words, in the context of realistic modelling of 
$\kappa$-type materials, the $d_{xy}$ symmetry found in the dimer model 
($d_{x^2-y^2}$ in the unfolded one-band model) is mostly an artifact of the 
underlying approximation to the real lattice structure 
(Eq.~\ref{eq:geometricformulas}).

Finally, we checked our results obtained with the four parameter molecule model 
against the original hopping structure obtained from projective Wannier 
functions, which includes longer range processes. As expected, the differences 
induced by the distance cutoff and parameter averaging are negligible.

\subsection{Pairing symmetry phase diagram of the molecule model}
To complete our study of the pairing symmetry competition, we investigated the 
leading pairing symmetry of the molecule model as a function of $t_2 / t_1$, 
$t_3 / t_1$ and $t_4 / t_1$ in the range of parameters realized in actual 
superconducting $\kappa$-type materials. 

\begin{figure}[tb]
\includegraphics[width=\linewidth]{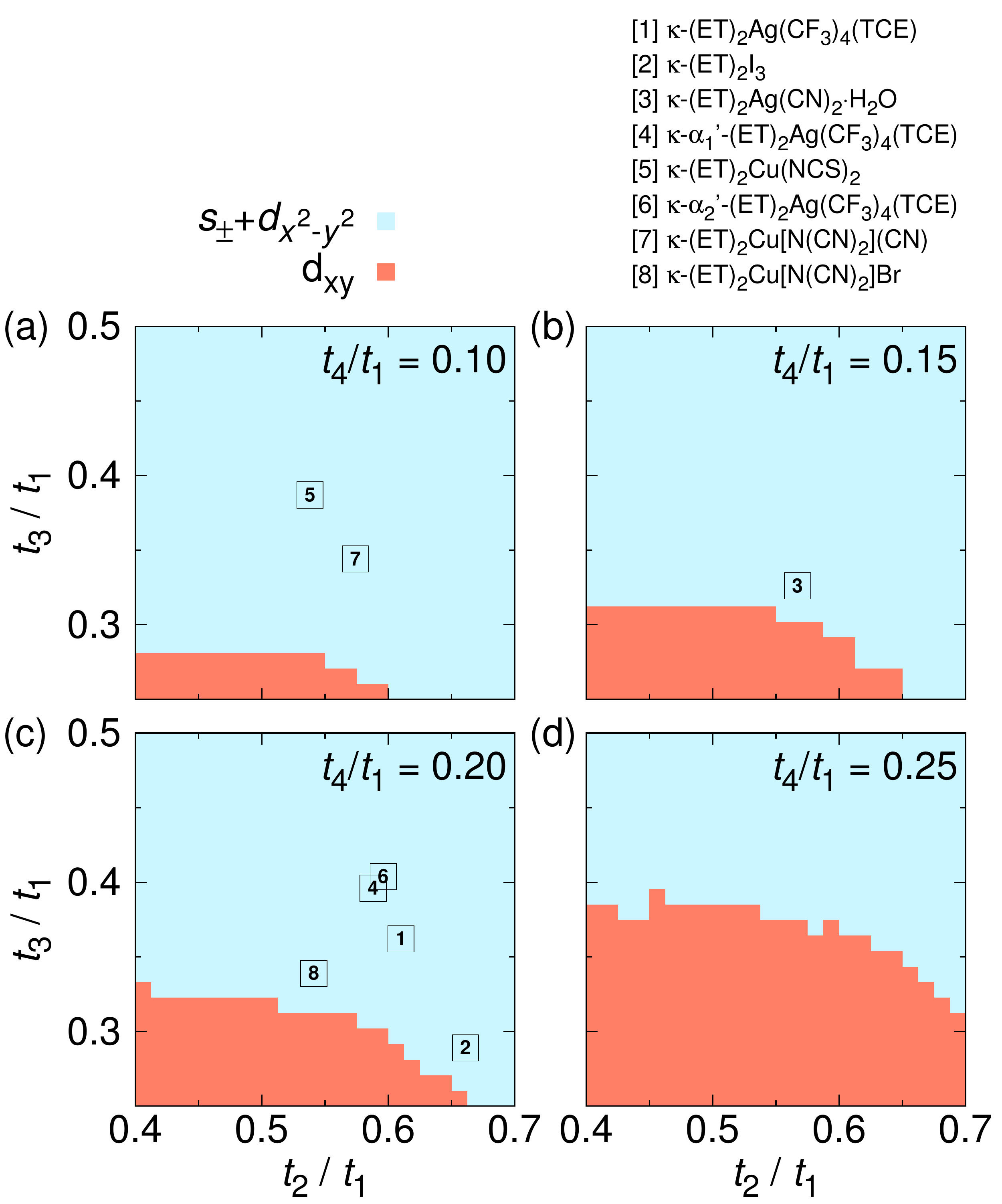}
\caption{(Color online) Superconducting phase diagram of the individual molecule 
model. Different symmetries of the superconducting order parameter are color 
coded. A d$_{xy}$ symmetry of the pairing interaction is favored when the 
orthorhombicity of the system is small, i.e. when $t_4 \lesssim t_2 $ and $t_3 
\ll t_2$. In the rest of the phase diagram an extended $\mathrm{s} + 
\mathrm{d}_{x^2-y^2}$ symmetry prevails. The numbered symbols correspond to the 
location of real materials in the phase diagram, enumerated as in 
Table~\ref{tab:materialparameters}. Materials were sorted into the subplot to 
which their true value of $t_4/t_1$ is closest.}
\label{fig:phasediagmolecule}
\end{figure}

In Fig.~\ref{fig:phasediagmolecule} we show the obtained phase diagram, which 
consists of a $d_{xy}$-symmetric phase at low $t_2 /t_1$ and $t_3 / t_1$, while 
the rest of the phase diagram shows a $s_\pm + d_{x^2-y^2}$ state. The 
consecutively numbered symbols in Fig.~\ref{fig:phasediagmolecule} correspond to 
the position of real materials as listed in Table~\ref{tab:materialparameters} 
within this phase diagram. As we scanned the phase diagram several times for 
different fixed $t_4 / t_1$, materials were sorted into the cut with the closest 
value of $t_4 / t_1$.

At low $t_2 /t_1$ the phase boundary is almost horizontal, i.e. independent of 
the precise value of $t_2 / t_1$. For larger values of $t_2 / t_1$ the model 
becomes more asymmetric with respect to $t_2$ and $t_4$ and a smaller diagonal 
coupling $t_3$ is sufficient to drive the system into the $s_\pm + d_{x^2-y^2}$ 
state. The size of the $d_{xy}$-symmetric region is obviously determined by the 
value of $t_4 / t_1$ as explained in the previous section.

\begin{figure}[t]
\includegraphics[width=\linewidth]{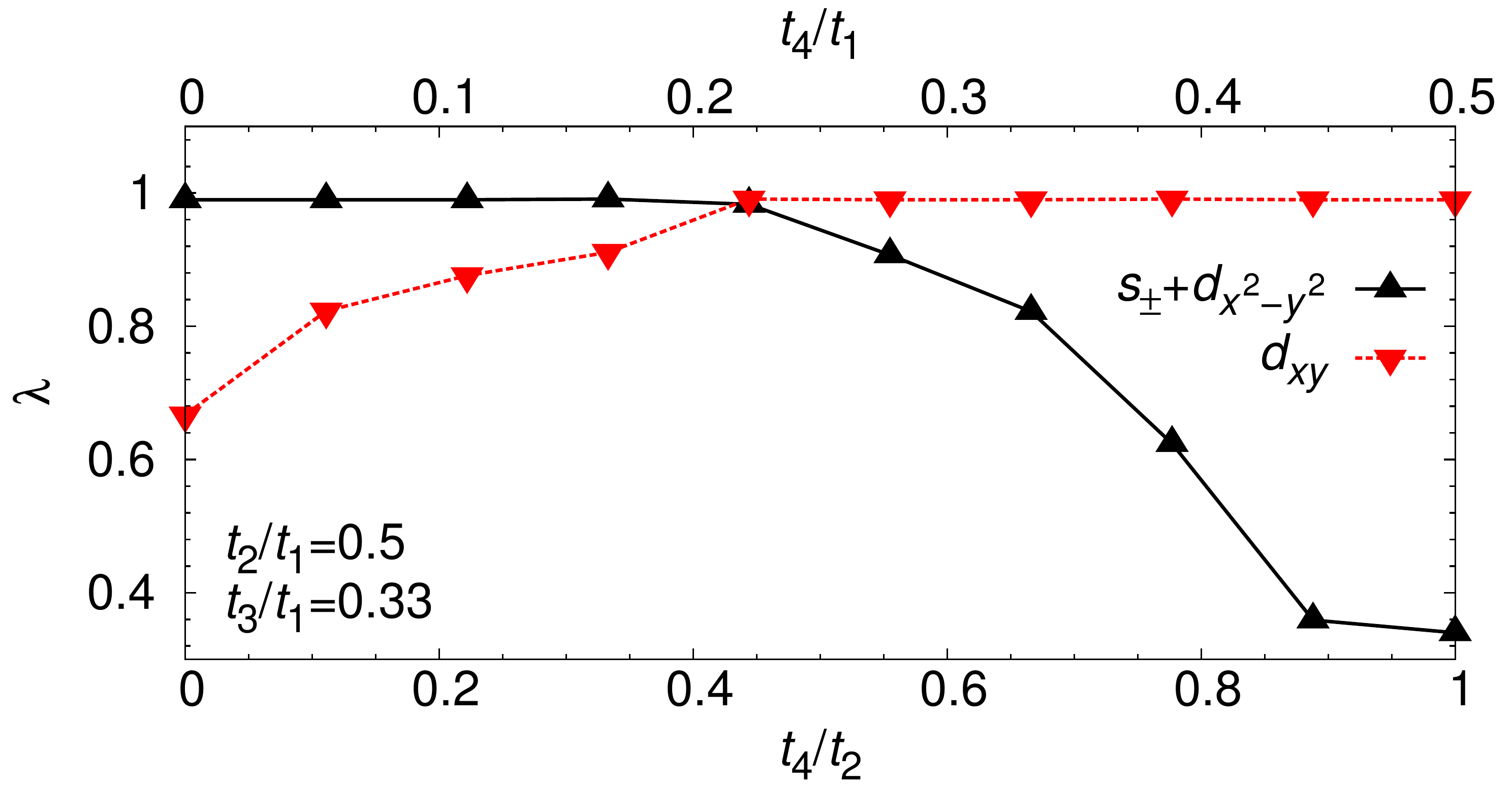}
\caption{(Color online) Eigenvalues of the gap equation for $s_\pm + 
d_{x^2-y^2}$ and $d_{xy}$ pairing symmetries as a function of the in-plane 
anisotropy $t_4/t_2$. The isotropic case is realized for $t_4/t_2 = 1$. Only 
$t_4$ was varied. The other parameters were fixed to $t_2/t_1 = 0.5$ and $t_3 / 
t_1 = 0.33$.}
\label{fig:symmetryeigenval}
\end{figure}

In the numerical calculations we observed that the leading two pairing 
symmetries are almost degenerate in a broad parameter region. This is to be 
expected, because the $s_\pm + d_{x^2-y^2}$ state emerges precisely as a 
compromise between two different nesting vectors, of which one rather fits to a 
pure $d_{xy}$-symmetry. To clarify this degeneracy, we calculated the 
eigenvalues of the leading and sub-leading solutions of the gap equation at 
fixed $t_2 / t_1$ and $t_3 / t_1$ and varied $t_4 / t_2$ in the range [0,1]. 
Fig.~\ref{fig:symmetryeigenval} shows the eigenvalues of both possible pairing 
states as a function of the in-plane anisotropy $t_4 / t_2$. We observe a 
pronounced asymmetry: While the $d_{xy}$-state is competitive even for low 
values of $t_4 / t_2$, the $s_\pm + d_{x^2-y^2}$ quickly becomes irrelevant when 
approaching the isotropic case ($t_4 / t_2 = 1$).

Finally, based on our parameter estimates, all materials investigated lie in the 
$s_\pm + d_{x^2-y^2}$ region of the phase diagram. Materials particularly close 
to the phase transition line are $\kappa$-(ET)$_2$I$_3$, 
$\kappa$-(ET)$_2$Ag(CN)$_2 \cdot $H$_2$O and $\kappa$-(ET)$_2$Cu[N(CN)$_2$]Br. 
These can be expected to realize the $s_\pm + d_{x^2-y^2}$ order parameter with 
eight nodes. Evidence for eight node mixed-symmetry superconductivity has recently
been found in Ref.~\onlinecite{Diehl2016}.

When materials are close to the phase transition line, small changes of the 
hopping parameters might drive them into the $d_{xy}$ state, which is always 
present as a sub-dominant pairing symmetry. For such local changes of 
parameters, for instance lattice defects~\cite{Analytis2006, Sano2010} or disorder of 
molecular conformations could be responsible. In 
Ref.~\onlinecite{EndgroupDisorder} we have shown that different conformations of 
ET molecules result in decidedly different ratios of $t_4 / t_2$. The degree of 
conformational disorder can be controlled experimentally by adjusting the sample 
cooling rate~\cite{Hartmann2014, Mueller2015}.

\begin{figure*}[t]
\includegraphics[width=\linewidth]{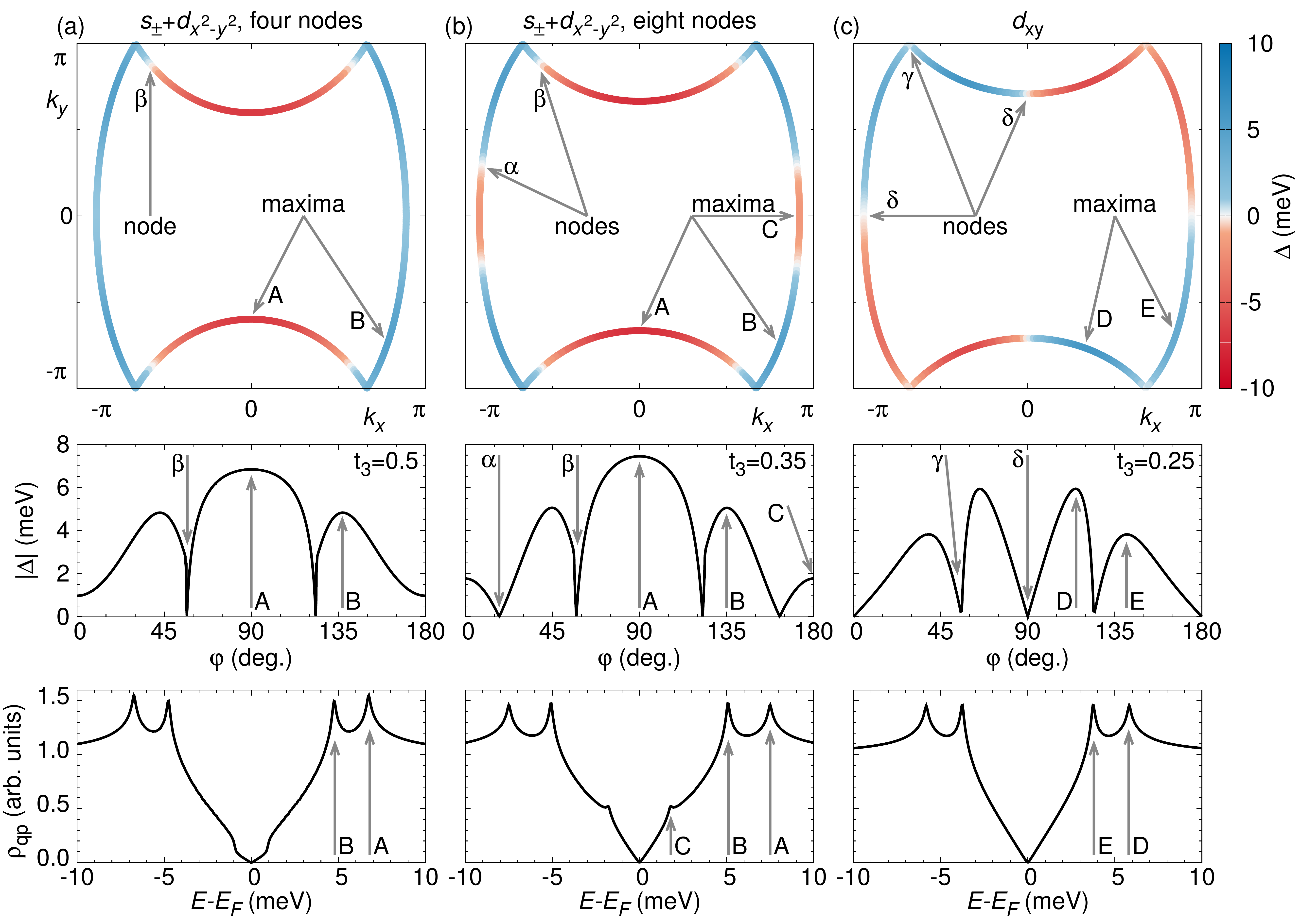}
\caption{(Color online) Gap function on the Fermi surface (top panel), magnitude 
of the gap function versus angle measured with respect to the $k_x$ direction 
(mid panel) and simulated quasiparticle density of states in the superconducting 
state (bottom panel). In all cases we assumed an energy scale $\Delta_0 = 
10~\mathrm{meV}$. Only $t_3 / t_1$ is varied. Other parameters are fixed to 
$t_2/t_1 = 0.4375$ and $t_4/t_2 = 0.1$. Maxima of the superconducting gap 
magnitude are labelled with capital letters. Nodes of the superconducting order 
parameter are labelled with greek letters. All nodes and maxima not labelled 
explicitly are symmetry equivalent to the labelled ones. Column (a) shows the 
case of $s_\pm + d_{x^2-y^2}$-symmetry with four nodes ($t_3 / t_1 = 0.5$). 
Column (b) shows the results for $s_\pm + d_{x^2-y^2}$-symmetry with eight nodes 
($t_3 / t_1 = 0.3475$). Column (c) shows the $d_{xy}$ case ($t_3 / t_1 = 
0.25$).}
\label{fig:stspectra}
\end{figure*}

In Ref.~\onlinecite{Kino1996} a square lattice-like antiferromagnetic order was 
found for the insulating state of $\kappa$-type materials. Therefore, we expect 
significant competition between antiferromagnetism and $d_{xy}$-symmetric 
superconductivity, while the $s + d_{x^2-y^2}$-symmetric state is realized 
farther away from the magnetically ordered insulating phase. Within this 
picture, recent results by Oka {\it et al.}~\cite{Oka2015}, who interpreted 
their experiment in terms of patches with a $d_{xy}$ order parameter and a 
$d_{x^2-y^2}$-symmetric bulk, can be qualitatively explained.

At this point we would like to point out that most experimental studies assume a 
four-node $d$-wave order parameter upon data analysis, which excludes from the 
start the detection of the $s_\pm$-component we found. In particular, the 
realization of the $s_\pm + d_{x^2-y^2}$ state with eight nodes which lie along 
the diagonals and close to the crystallographic axes, may  explain the 
considerable disagreement in the experimental literature regarding the node 
positions.

\subsection{Simulation of scanning tunneling spectroscopy}
Most transport experiments on $\kappa$-(ET)$_2 X$ materials have proven to be 
difficult to interpret and could not resolve the symmetry of the superconducting 
pairing so far. However, recent improvements in sample preparation for 
low-temperature scanning tunneling spectroscopy (STS) experiments have allowed 
for progress towards a resolution of the superconducting order 
parameter~\cite{Oka2015, Diehl2014, Diehl2016}.

Therefore, in this section we simulate tunneling spectra in the superconducting state for 
molecule model parameters $t_2/t_1 = 0.4375$, $t_4/t_1 = 0.1$ and various values 
of $t_3 / t_1$: a four-node $s_\pm + d_{x^2-y^2}$ state is obtained for $t_3 / 
t_1 = 0.5$, an eight-node $s_\pm + d_{x^2-y^2}$ state for $t_3 / t_1 = 0.3475$ 
and $d_{xy}$ for $t_3 / t_1 = 0.25$.
We employ the representation of the superconducting gap in terms of symmetry 
functions introduced in Eq.~\ref{eq:modelgap}, which we multiply with a 
prefactor $\Delta_0 = 10~\mathrm{meV}$ to obtain a spectrum with reasonable 
energy scale. The gap on the Fermi surface is then given by $\Delta (\vec k) = 
\Delta_0 \, \tilde g (\vec k)$. We use this expression together with 
Eq.~\ref{eq:dosqpfinal} to calculate the quasiparticle density of states 
$\rho_{\textnormal{qp}}$, which corresponds to the local density of states (LDOS) observed in 
STS experiments. The finite quasiparticle lifetime is modelled by $\Gamma = 
0.07~\mathrm{meV}$. In the $d_{xy}$ case we ignore the small anisotropy found in 
the RPA calculation. 

In Fig.~\ref{fig:stspectra} we show (i) the obtained gap on the Fermi surface, 
(ii) the magnitude of the gap versus angle measured from the $k_x$-direction and 
(iii) the simulated tunneling spectrum for the three cases investigated.

The magnitude of the gap versus the angle is distributed anisotropically on the 
Fermi surface [Fig.~\ref{fig:stspectra}(a-c) top panel]. Maxima of the gap 
magnitude are indicated by arrows labelled with capital letters, while nodes in 
the gap are indicated by arrows labelled with greek letters. The global maximum 
of the gap magnitude (labelled A or D) resides in all cases on the elliptic part 
of the Fermi surface, while the second largest maximum (labelled B or E) is 
located on the quasi-one-dimensional part. A third smallest maximum (labelled C) 
is possible on the quasi-1D sheet. In the $d_{xy}$-case the nodes labelled 
$\gamma$ appear in addition to the expected set of nodes $\delta$, because the 
Fermi surface touches the Brillouin zone boundary. The $\gamma$-nodes lead to 
the second maximum (labelled E) of the gap magnitude 
[Fig.~\ref{fig:stspectra}(a-c) middle panel], but are otherwise irrelevant for 
the low-energy physics. As the three possible gap structures share two maxima of 
slightly different size, the simulated quasiparticle DOS looks quite generic 
[Fig.~\ref{fig:stspectra}(a-c) bottom panel]. A two-peak structure is observed 
far away from the Fermi level, which corresponds to the energy values of the two 
largest maxima in the gap magnitude.

Important differences are however revealed at low energies: the $d_{xy}$-state 
is featurelessly V-shaped [Fig.~\ref{fig:stspectra}(c) bottom panel], while the 
spectrum of the eight-node state has an additional peak close to 2~meV 
[Fig.~\ref{fig:stspectra}(b) bottom panel], which is linked to the small gap 
(labelled C) on the quasi-1D sheet. This leads to an outer and an inner V-shape 
with different slopes. For the four-node $s_\pm + d_{x^2-y^2}$-state we observe 
a peculiar dip around 1~meV in the quasiparticle spectrum 
[Fig.~\ref{fig:stspectra}(a) bottom panel]. This corresponds to the minimum 
value of the gap magnitude on the quasi-one-dimensional part of the Fermi 
surface. Inside of this dip a V-shaped region emerging from the $\beta$-nodes is 
again observed.

We emphasize that our predictions are to be taken as qualitative, not 
quantitative, regarding the overall energy scale and the relative gap sizes. The 
main features explained above are however robust. The detection of such 
low-energy structures is certainly not an easy task, but we believe it will be 
possible with state-of-the-art equipment and proper sample preparation.

\section{Conclusions}
In summary, we investigated the superconducting state of $\kappa$-(ET)$_2 X$ 
charge transfer salts in an individual molecule model based on a combination of 
{\it ab initio} density functional theory and random phase approximation 
spin-fluctuation calculations. We obtained kinetic parameters of the molecule 
Hamiltonian for eight superconducting $\kappa$-type materials using projective 
Wannier functions. We found that the superconducting order parameter in a 
realistic molecule model is different from the one in the usual dimer 
approximated Hamiltonian for all investigated materials. The superconducting 
phase diagram of the molecule description is dominated by an extended $s + 
d_{x^2-y^2}$-symmetry that emerges from the competition between square-like and 
diagonal hopping processes on the original $\kappa$-type lattice, while the 
physics of the dimer model is reproduced also for finite dimerization in the 
limit of isotropic parameters $t_4 \lesssim t_2$. The anisotropy of square-like 
hoppings $t_2$ and $t_4$ is however not negligible in real materials. For 
precisely this reason, the dimer approximation does not apply to superconducting 
$\kappa$-(ET)$_2 X$ charge transfer salts. It overestimates the importance of 
square lattice physics through the averaging contained in the geometric 
formulas, which are exact only in the limit of infinite dimerization. 

Furthermore, the $s_\pm + d_{x^2-y^2}$-state, which features nodes both along 
the crystallographic axes and the Brillouin zone diagonals, might explain the 
multitude of contradictory experimental results regarding the nodal positions. 
We also simulated tunneling spectroscopy experiments for all nodal 
configurations encountered in our phase diagram. The difference between those  
pairing states unfortunately manifests itself only at very low energies, making 
experimental detection difficult, but not impossible.
Based on the {\it ab initio} calculated model parameters we found that the 
well-studied material $\kappa$-(ET)$_2$Cu[N(CN)$_2$]Br is situated near the 
phase transition line between $s_\pm + d_{x^2-y^2}$ and $d_{xy}$ superconducting 
states, which supports the interpretation of recent scanning tunneling 
spectroscopy experiments.

A question unanswered by our study is why superconducting transition 
temperatures among quasi-two-dimensional charge transfer salts can differ by 
more than a factor of four. As there is no obvious connection between $T_c$ and 
the parameters of the kinetic Hamiltonian, a method that can qualitatively 
reproduce the ordering of transition temperatures in real materials is required 
to elucidate this issue.

In conclusion we believe that a significant part of the physics in 
quasi-two-dimensional charge transfer salts has unfortunately been 
overlooked so far, because 
theory has adhered to the dimer model for too long and too many experiments have 
been interpreted based on a dichotomy of $d_{xy}$ and $d_{x^2-y^2}$ states, 
which is inappropriate for the orthorhombic lattice realized in $\kappa$-(ET)$_2 
X$ materials. 

It is an interesting open question, whether the magnetic, insulating and possible quantum
spin-liquid states known from the anisotropic triangular lattice are also 
present in the molecule model. The investigation of these phases is left for future studies.

\begin{acknowledgments}
The authors acknowledge fruitful discussions with Ryui Kaneko, Stephen M. 
Winter, Andreas Kreisel and Peter J. Hirschfeld. This work was supported by the 
German Research Foundation (Deutsche Forschungsgemeinschaft) under grant SFB/TR 
49. Calculations were performed on the LOEWE-CSC and FUCHS supercomputers of the 
Center for Scientific Computing (CSC) in Frankfurt am Main, Germany.
\end{acknowledgments}

\appendix

\section{Matrix elements of the kinetic Hamiltonians}
\label{sec:kinhamil}
For completeness we list the kinetic part of the dimer model in one- and 
two-band representation, as well as the kinetic part of the four-band molecule 
model. We denote the unit cell parameters in $x$- and $y$-direction as $a$ and 
$b$ respectively. The multi-band Hamiltonians are given as matrix elements 
$\bra{i}H_\text{hop}\ket{j}$, where states $\ket{i}$ denote the orbitals living 
on a dimer/molecule with site index $i$. Only unique matrix elements are listed. 
The rest of the elements are generated by using $\bra{i}H_\text{hop}\ket{j} = 
\bra{j}H_\text{hop}\ket{i}^*$. In all models there is only one orbital per 
lattice site. 

To obtain the correct electron filling, one has to introduce a chemical 
potential $\mu$, so that $H_0 = H_\text{hop} - \mu\sum_{i 
\sigma}c_{i\sigma}^\dagger c_{i \sigma}$ is half-filled for the dimer model in 
either representation and 3/4-filled for the molecule model.

The single-band representation of the dimer model is given by:
\begin{equation}
\begin{array}{rl}
H_\text{hop}(\vec k) =& 2t \left[ \text{cos} (k_x a) + \text{cos} (k_y b) 
\right] \\[4pt]
&+2t^\prime \left[\text{cos}(k_x a)\,\text{cos}(k_y b) - \text{sin}(k_x 
a)\,\text{sin} (k_y b) \right]
\end{array}
\label{eq:singlebanddimerhamil}
\end{equation}
The two-band representation of the dimer model can be written as:
\begin{subequations}
\begin{align}
\bra{0}H_\text{hop}\ket{0} &= \bra{1}H_\text{hop}\ket{1} = 2 t^\prime \text{cos} 
(k_x a) \\
\bra{0}H_\text{hop}\ket{1} &= 2t\left( 1 + e^{i k_x a} + e^{i k_y b} + e^{i k_x 
a} e^{i k_y b} \right)
\end{align}
\label{eq:twobanddimerhamil}
\end{subequations}
The four-band molecule model is given by:
\begin{subequations}
\begin{align}
\bra{0}H_\text{hop}\ket{1} &= t_1 + t_3 \, e^{i k_x a} \\
\bra{0}H_\text{hop}\ket{2} &= t_4 \left( 1 + e^{-i k_y b} \right) \\
\bra{0}H_\text{hop}\ket{3} &= t_2 \left( 1 + e^{-i k_x a} \right)  \\
\bra{1}H_\text{hop}\ket{2} &= t_2 \, e^{-i k_y b } \left( 1 + e^{-i k_x a} 
\right)  \\
\bra{1}H_\text{hop}\ket{3} &= t_4 \, e^{-ik_x a} \left( 1 + e^{-i k_y b} \right) 
\\
\bra{2}H_\text{hop}\ket{3} &= t_1 + t_3 \, e^{-i k_x a}
\end{align}
\label{eq:fourbandmoleculehamil}
\end{subequations}
Here, diagonal entries $\bra{i}H_\text{hop}\ket{i}$ are zero for all $i$. 


\bibliographystyle{apsrev4-1}

\end{document}